\documentclass[usenatbib, useAMS, usegraphicx]{mn2e}

\begin{document}

\title[2D K-S Test for Source Detection]
	{A two-dimensional Kolmogorov--Smirnov test for crowded field 
	source detection: {\it ROSAT} sources in NGC 6397}

\author[S.~A.~Metchev and J.~E.~Grindlay]{S.~A.~Metchev$^1$ and
J.~E.~Grindlay$^2$\\ 
$^1$Department of Astronomy, California Institute of Technology, 
Pasadena, CA 91125, USA\\
$^2$Harvard-Smithsonian Centre for Astrophysics, 60 Garden Street, 
Cambridge, MA 02138, USA}

\maketitle

\begin{abstract}
We present a two-dimensional version of the classical one-dimensional
Kolmogorov--Smirnov (K-S) test, extending an earlier idea due to
\citet{pea83} and an implementation proposed by \citet{fas87}.
The two-dimensional K-S test is used to optimise the goodness of fit in
an iterative source-detection scheme for astronomical images.
The method is applied to a {\it ROSAT}/HRI x-ray image of
the post core-collapse globular cluster NGC~6397  to determine the
most probable source distribution in the cluster core.  Comparisons to
other widely-used source detection methods, and to a {\it Chandra}
image of the same field, show that our iteration scheme is superior in
measuring statistics-limited sources in severely crowded fields.
\end{abstract}

\section{Introduction}
Deep x-ray imaging of crowded fields, even with  increasing angular
resolution and sensitivity (with {\it Einstein} and {\it ROSAT}, and now
{\it Chandra}), is invariably limited by the small number of source counts
and by the relative size of the
point-spread function (PSF) compared to the angular separation between the
objects.  Determining the underlying source configuration in such a
regime is  often beyond the capabilities of conventional
source-finding algorithms.

Classical x-ray source detection methods are based on a sliding detection 
cell of a fixed size across the image, and calculating the
signal-to-noise ratio ($S/N$) at each step.  To find $S/N$, common 
detection algorithms for processing data from
{\it Einstein} and {\it ROSAT} (implemented in {\sc
iraf}\footnote{Image Reduction 
and Analysis Facility; developed and maintained by the National
Optical Astronomy Observatories.}{\sc /pros}\footnote{Post-Reduction
Off-line Software; developed and maintained by the Smithsonian
Astrophysical Observatory.}) use either an average background (as
determined from a source-free section of the image) or a local
background (from a region around the detection cell). However, both
methods fail to discern blended faint sources in crowded fields
where the background is affected by overlapping PSFs.  Source detection 
is somewhat improved by image
deconvolution, e.g.\ with the Lucy--Richardson (L-R) algorithm or with
the Maximum Entropy Method (MEM), or by wavelet smoothing.   Deconvolution
algorithms provide higher positional sensitivity in moderately crowded
fields, but suffer from such undesirable effects as
noise-amplification and ``leakage'' (associating counts from fainter
sources to brighter nearby ones). Wavelet detection  
implemented as task {\sc wavdetect} in the {\it Chandra} processing
package (available  
at http://asc.harvard.edu/ciao) does well in crowded fields, provided 
the sources are either sufficiently separated ($\ga$3--5~FWHM, or 
$\ga$3--5~arcsec) or within $\sim$2--3~FWHM {\it and} are similar in
flux \citep{dam97, fre01}.
When the PSFs are heavily  blended (separation between the source
centroids $\la 1.5$~FWHM), 
individual sources cannot be distinguished and their  relative fluxes
cannot be measured.  A superior source-detection method is 
needed for severely crowded fields containing multiple
faint sources, e.g.\ globular cluster cores \citep{her83}, or
nuclear bulges in  external galaxies.

A  powerful technique to compare statistics-limited
samples is the Kolmogorov--Smirnov (K-S) test which, unlike its
alternative -- the
Pearson $\chi^2$ test -- does not require binning of the
data. Unfortunately, the classical K-S test is applicable only to
one-dimensional distributions, and any attempts to convert a
two-dimensional image to one dimension (e.g.\ by collapsing it onto a
vector, or by azimuthal binning around a point) lead to unwanted loss
of information and power.  For some time now, a multi-dimensional
version of the K-S test has been known \citep{pea83, gos87}, which performs
better than the $\chi^2$ test in the small-number statistics
case \citep[][hereafter, FF]{fas87}, and can be successfully applied in
parameter point estimation in a manner similar to the widely used
maximum-likelihood (ML) method.  These properties of the
multi-dimensional K-S test make it viable for incorporation in
source-detection methods.

In this paper, we re-visit the characteristics of the K-S test in two
(and three) dimensions and examine its power in comparing different
realisations of crowded low $S/N$ fields.  As an application of the
test, we devise an iterative source-modelling scheme that aims to minimise the
K-S statistic in search of the optimum underlying source distribution in an
image.  Based on our Monte Carlo simulations, we find that our
iterative algorithm is a powerful tool for faint object searches in
crowded fields.  We apply the algorithm to determine the faint x-ray
source distribution in a deep {\it ROSAT} exposure of the post core-collapse
globular cluster NGC~6397, which has also been analysed with ML
techniques by \citet{ver00}. We compare the derived
x-ray positions with those of \citeauthor{ver00} and with our subsequent
optical \citep[{\it HST}, ][]{tay01} and x-ray \citep[{\it
Chandra}, ][]{gri01a} identifications.

\section{The two-dimensional K-S test}

\subsection{Description \label{sec_description}}

The classical one-dimensional (1D) K-S test makes use of the
probability distribution of the quantity $D_{KS}$, defined as the
largest absolute difference between the cumulative frequency
distributions of the parent population and that of an $n$-point sample
extracted from it.  Since $D_{KS}$ is approximately proportional to
$1/\sqrt{n}$, one usually refers to the probability distribution of
the quantity $Z_n \equiv D_{KS}\sqrt{n}$.  For a given $n$, the values
of $Z_n$ corresponding to a given significance level $SL$ (denoted as
$Z_{n,SL}$) increase slightly with $n$.  For large $n$, the integral
probability distribution $P(>Z_n) = 1-SL$ approaches the  asymptotic
expression \citep{ken79}:
\begin{displaymath}
P(>Z_n)=2 \sum_{k=1}^{\infty} (-1)^{k-1} \exp(-2k^2Z_n)
\end{displaymath}
which is satisfactory for $n \geq 80$.  For the two-sample K-S test,
which compares distributions of different sizes ($n_1$ and $n_2$),
the probability
distribution $P(>Z_n)$ remains unchanged provided that $n$ is set to
$\frac{n_1n_2}{n_1+n_2}$.  The 1D nature of the test implies that it
does not depend in any way on the shape of the parent distribution.

In a two-dimensional (2D) distribution, each data point is
characterised by a pair of values, $(x,y)$.   As with the 1D K-S test,
the maximum cumulative difference between two 2D distributions is
found over the $(x,y)$-plane.  In the case of
distributions in more  than one dimension however, the  procedure to cumulate
the information onto the plane is not unique.  FF
made use of the total number of points in each of the four
quadrants around a given point $(x_i,y_i)$, namely, the fraction of data points
in the regions $(x<x_i, y<y_i), (x<x_i, y>y_i), (x>x_i,
y<y_i), (x>x_i, y>y_i)$.  The 2D statistic $D_{KS}$ is defined as the
maximum difference between data fractions in any two matching quadrants of 
the sample 
and of the parent population, ranging over all data points.  The $Z_n$
statistic is defined similarly as in the 1D case, $Z_n \equiv
D_{KS}\sqrt{n}$, where for a two-sample 2D K-S test $n \equiv
\frac{n_1 n_2}{n_1+n_2}$.

Based on their Monte Carlo simulations, FF deduce that the
2D integral probability distribution $P(>Z_n)$ depends
solely on the correlation coefficient ($CC$) of the model
distribution, i.e.\ that for a given $CC$ the distribution of $Z_n$ in 
the 2D K-S test is (nearly) independent of the shape of the model, as in 
the classical 1D K-S test. FF also
observe that in the two-sample case, it is sufficient to take the
average of the correlation coefficients $CC_1$ and $CC_2$ of the
samples as an estimate of $CC$.

An important distinction between \citeauthor{pea83}'s and FF's version
of the 2D K-S test was pointed out to us by the referee, which makes the
latter generally less stringent.  FF
restrict the search for the maximum cumulative difference $D_{KS}$ to
loci harbouring a data point, thereby often missing the location of
the true maximum difference, which is almost always found for $(x<x_i, y<y_j)$,
where $i$ and $j$ are two different data points.  Nevertheless, the
maximum cumulative difference computed in such a way will have a
tendency to vary in the same manner as the true maximum difference.  Thus
the FF statistic is probably well-behaved, at least
as long as the genuine parent population distribution and the assumed
one are not too different \citep{gos87}.  The latter situation is indeed
expected when comparing distributions of point-spread functions in
two images.  The advantage of FF's approach is speed: order $n$, instead
of $n^2$.  The disadvantage is its approximate nature and that the
$D_{KS}$ statistic is sensitive to the
correlation coefficient $CC$ of the distributions, requiring its
inclusion as a free parameter in the reference tables.  Our Monte Carlo
experiments below take into account both factors.
 
\subsection{Monte Carlo experiments}

Following the procedure in FF, we used the 2D K-S test computer code
provided in {\sl Numerical Recipes} \citep{pre97} to run our own
Monte Carlo experiments.  We studied the $Z_n$ statistic by means
of a Monte Carlo procedure using a uniform distribution ($CC=0$) within 
the unit square as a parent population.  The analysis comprised of cases with
number of points $n$ per sample ranging from
$n=5$ to $n=50000$.  For any given $n$ we produced a large number
of simulations (from 100000 for $n=5$ to 1000 for $n=50000$), 
enabling us to construct the integral probability distribution
$P(>Z_n)$ with sufficient accuracy. Values of $Z_{n,SL}$ for uniform 
samples of all tested sample sizes $n$ are listed in Table~\ref{tbl_cc0.0}.

\begin{table*}
\begin{minipage}{115mm}
\caption{Critical values $Z_{n,SL}$ for a uniform uncorrelated
distribution.} 
\label{tbl_cc0.0}
\begin{tabular}{@{}rrccccccccc}
 & $SL(\%)$$^\dag$ & 30 & 40 & 50 & 60 & 70 & 80 & 90 &
95 & 99 \\ 
$n$$^\ddag$ & \# of simul. & & & & & & & & & \\
5&100000&0.76&0.82&0.88&0.95&1.03&1.13&1.27&1.38&1.62 \\
10&100000&0.79&0.85&0.92&0.98&1.06&1.15&1.29&1.41&1.65 \\
20&100000&0.85&0.91&0.97&1.03&1.10&1.20&1.33&1.45&1.69 \\
50&100000&0.91&0.97&1.04&1.10&1.18&1.27&1.41&1.53&1.78 \\ 
100&10000&0.96&1.02&1.08&1.15&1.22&1.32&1.45&1.57&1.81 \\ 
200&5000&1.01&1.07&1.13&1.19&1.27&1.37&1.51&1.62&1.86 \\ 
500&5000&1.06&1.12&1.18&1.25&1.33&1.43&1.57&1.68&1.91 \\ 
1000&5000&1.09&1.15&1.21&1.28&1.35&1.45&1.60&1.73&1.98 \\ 
2000&5000&1.11&1.17&1.24&1.30&1.38&1.47&1.61&1.74&1.99 \\ 
5000&2000&1.14&1.21&1.27&1.34&1.42&1.50&1.64&1.74&1.96 \\ 
10000&1000&1.15&1.22&1.28&1.35&1.43&1.54&1.69&1.82&2.13 \\ 
20000&1000&1.14&1.21&1.29&1.37&1.45&1.56&1.74&1.89&2.20 \\ 
50000&1000&1.16&1.24&1.32&1.40&1.50&1.60&1.78&1.91&2.19 \\ 
\end{tabular}

\medskip
$^\dag$Significance level ($SL \equiv 1-P(>Z_n)$).

$^\ddag$Size of sample.
\end{minipage}
\end{table*}

\begin{figure}
\includegraphics{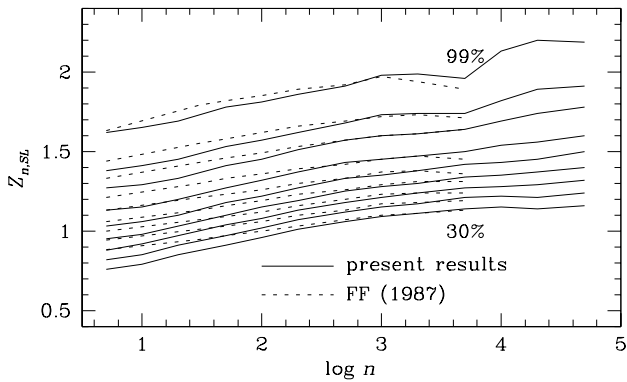}
%\vspace{5.5cm}
\caption{Critical values of the statistic $Z_n$ as a
function of sample size $n$, for values of the significance level $SL$
varying from 30 per cent to 99 per cent.  The continuous lines are
based on data listed in Table~\ref{tbl_cc0.0}, and the dashed lines
are based on data reported in Table~A1 of FF.} 
\label{fig_zn_logn}
\end{figure}

Figure~\ref{fig_zn_logn} presents a comparison between our and FF's
results for the critical values $Z_{n,SL}$ as a function of $n$.  Both
sets of Monte Carlo simulations show similar tendencies in the
behaviour of the $Z_n$ statistic, and are indistinguishable from each
other for $n \geq 500$ within the statistical uncertainties ($\approx
5$~per cent for high $n$, due to the limited number of simulations).  There
is however a marked inconsistency between the two sets of data in the
low-$SL$, small-$n$ part of the graph, where at the 30 per cent significance
level, the values of $Z_{n,SL}$ differ by a factor of $\sim 1.15$.
The difference is highly significant, given the fact that both FF's
and our results for small $n$ are based on 100000 simulations.  We
attribute this discrepancy to a detail in the implementation of the 2D
K-S test: in particular, to whether the data point around which the
$D_{KS}$ statistic is computed, is included in one of the cumulative
quadrants or not.  This discrepancy becomes irrelevant, when the same
implementation of the 2D K-S test is used in both building the
reference tables for the critical values of $Z_n$, and in comparing
actual distributions using these tables.

\section{Using the two-dimensional K-S test on astronomical images}

FF applied a 2D K-S test to  astronomical
distributions (not images) and showed that it can reject wrong hypotheses
at a much higher significance level than the $\chi^2$ test.  We extend
the application of the 2D K-S test to comparing images of crowded fields, 
where it is of great
interest to determine whether a proposed source distribution
corresponds to the observed one.  When the 
PSFs of the individual sources are heavily blended (source separation
$\la 1.5$~FWHM), classical 
x-ray source-detection methods fail to distinguish the individual objects.
Crowded-field optical photometry tools \citep[e.g.\ {\sc daophot},][]{ste87,
ste91} are also not suitable for the small-number Poisson statistics of
x-ray images.  Even the recently introduced detection algorithm
based on wavelet transforms \citep[][and references therein]{fre01, dam97} 
implemented in the {\it Chandra} x-ray data analysis package 
({\sc CIAO}\footnote{{\it Chandra} Interactive Analysis and
Observations; developed and maintained by the {\it Chandra} X-ray Center, 
and available at http://asc.harvard.edu/ciao.}),
does not produce adequate results in the regime of severe source
confusion and small number of counts per source.

Below we describe an implementation of the 2D K-S test to
astronomical images.  Provided that the PSF of the (unresolved)
sources in a crowded field is known and constant over the image, by
comparing the image to a simulation of a proposed source
distribution, we can obtain the K-S probability $P(>Z_n)$ that the
image and the simulation represent the same source configuration.  The
obtained probability can be used as a measure of the accuracy of both
the positions and the intensities of the proposed sources, as well as
an indication of the necessity for additional sources to account for
the photon distribution.  Because of the high sensitivity of the 2D
K-S test, we expect that it should be able to discern positional
discrepancies equal to a fraction of the FWHM of the PSF.

\subsection{Implementation of the two-dimensional K-S test}

\subsubsection{A two-dimensional vs.\ a three-dimensional K-S test
\label{sec_2d3d}}

Astronomical images (e.g.\ from CCDs) have three dimensions:
$x$ and $y$ pixel coordinates, and pixel intensity.  Although the
spatial distribution of photons that strike the detector is
two-dimensional (two photons never fall at the exact same position),
they are binned by the detector in integer bins corresponding to the
digital pixel size.  Thus the incoming 2D distribution of photons with
real-valued coordinates is transformed into a three-dimensional (3D)
one with integer coordinates.  It is possible to apply the 2D K-S test
to an integer-valued 3D distribution by simply re-calculating the
two-dimensional $D_{KS}$ statistic for every count in a
given pixel.  However, since the K-S test is designed to deal with
real-valued distributions, any point in the $(x,y)$ plane that
has more than one count at the {\sl exact} same real-valued position (which
would be the case if a pixel contained more than one count) becomes
disproportionately significant, and distorts the value of $D_{KS}$.

Therefore, to apply a 2D K-S test to 3D images, it is first
necessary to ``un-bin'' the images by spreading the counts in
each pixel over the area of the pixel.  Since no information is
preserved about the exact impact locations of the photons on the
detector, we introduce a random shift (between $-0.5$ and $+0.5$ pix)
in the integer-valued coordinates of each count.  Every
count in the image is thus assigned a unique position (within the precision
limits of the computer), and the integer-valued 3D coordinates 
are converted to real-valued 2D coordinates.
Since this routine distributes the data on a sub-pixel scale, we refer
to it as ``subpixelization.''

Unfortunately, subpixelization incurs an undesirable effect, due to
the randomness with which the counts are moved around within the
pixels.  In particular, two subpixelized versions of the same image
are never the same.  Therefore, K-S testing of different
subpixelizations of the same two images will produce different results
for $Z_n$ every time.  Our experience is that $Z_n$ varies by about
2 to 4~per cent between runs; the corresponding variations in the
significance level $SL$ may be as high as $\pm 8$~per cent for values of
$Z_n$ near $SL=50$~per cent (Table~\ref{tbl_cc0.0}). To obtain a mean value
for the K-S probability $P(>Z_n)$ (equal to $1-SL$) with which two
images represent the same parent distribution, it is necessary to run
the 2D K-S test multiple times. The error in the K-S probability can
be estimated from the variations in $Z_n$ among the different runs.

The above method may appear contrived and unnecessary, when instead of
applying a 2D K-S test, by following the generalisation in FF, a 3D
K-S test can be implemented.  A 3D test does not have the undesirable
uncertainties associated with subpixelization, and gives the exact
K-S probability that two images represent the same parent
distribution.  FF report that the 3D test exhibits greater power in
rejecting wrong hypotheses than a three-fold 2D test along each of the
$(x,y)$, $(y,z)$ and $(z,x)$ planes.  However, our experiments with
the tests on simulated {\it ROSAT} images show that the 2D K-S test applied
to subpixelized images is more powerful than its 3D counterpart (with
accordingly generated look-up tables) applied to the original (not 
subpixelized) images.
K-S tests performed on simulations of an 880-count five-source
distribution with  centre-to-centre distances of 1.1--2.1 times the
PSF size (5~arcsec for  {\it ROSAT}/HRI) show that the 2D test on subpixelized
images can distinguish individual source position shifts as small as
2--3~arcsec (depending on the relative source locations and intensity) at
the $\ga 97$~per cent significance level.  The 3D test finds such shifts
insignificant: it gives a probability  of $\ga 70$~per cent that the
simulations represent {\sl the same} parent  distribution.

The higher power of the subpixelized 2D test is explained
by the manner in which pixels are weighted. Subpixelization ensures
that all photons are assigned equal weights: a 
desirable effect, since pixels are weighted proportionally to their
intensity.  The 3D K-S test, on the other hand, assigns equal weights
to all pixels, so single-count pixels (often from background) are as
significant as pixels with multiple counts (denoting sources).
Hence, the 2D
test on subpixelized images is more sensitive to variations in the
source distribution than its 3D counterpart.

We also note that subpixelized images represent more realistically the
incoming photon distribution.  Repeated 2D K-S tests on independent
subpixelizations produce an estimate of the random error in the K-S
probability induced by photon binning.

\subsubsection{One-sample vs.\ two-sample two-dimensional K-S test
\label{sec_1s2s}}

When comparing an image to a proposed model distribution of sources
(using an analytical or a fitted PSF), perfect information is
available about the shape of the model distribution.  It is therefore
appropriate to use a one-sample K-S test to compare the image to
the model distribution.  However, the analytical form of a fitted PSF
is often very complex.  The added complication of having several nearby
sources with overlapping PSFs (as in a crowded field) makes it
computationally very tedious to calculate the fraction of the analytic
model distribution in the quadrants around every count in the image
(needed to compute the $D_{KS}$ statistic; Section~\ref{sec_description}).

To avoid lengthy 2D surface integrals, the two-sample K-S test can be
used instead, to compare the image to a simulation generated from the
proposed model.  The random deviations in the representation of the model
can be decreased if a ``bright'' (high number of counts
$n$) simulation is used, that follows the model closely.  The
fractional deviation of 
the simulated vs.\ expected counts per pixel will decrease as
$1/\surd{n_{\rm cts}}$ for large $n_{\rm cts}$, where $n_{\rm cts}$ is
the number of counts per pixel in the model image.  The
running time of the two-sample 2D K-S test is however proportional to the
square of the sum $n_1 + n_2$ of counts in the samples compared, and
it is impractical to use the two-sample K-S test on simulations
containing high number of  counts.

To take advantage of the higher power of the one-sample test (given
perfect information about the shape of the model distribution), and to
avoid long running time, we compare the image (containing $n$ counts)
to a bright simulation ($n_{\rm mod} = k \times
n$ counts; $k\sim 50$) of the 
proposed model, using the {\sl one-sample} test.  In essence, we use
Monte Carlo integration for the model, the assumption being that a
bright simulation can be made to represent the model with sufficient
accuracy, and simple summing of the counts in the quadrants can be
substituted for analytic integration of the PSF.  We therefore do not 
expect the properties of this {\sl pseudo} one-sample K-S test to be
significantly different from those of the FF one-sample K-S test.  In
particular, we assume that the new test is still distribution-free for
a fixed correlation coefficient $CC$ of the bright simulation
(hereafter referred to as the ``model''), and that its properties vary
slowly with $CC$ (as observed in FF for their 2D K-S test).
Clearly, our test will converge to the FF one-sample test for $k
\rightarrow \infty$, but will run faster than the latter for
moderate-sized $k$, since lengthy analytical integrations are substituted 
with Monte Carlo integration.  The pseudo one-sample test (running time 
proportional to $n(n+n_{\rm mod}) = (k+1)n^2$) is also $k+1$ times faster 
than the {\sl two}-sample test (running time
proportional to $(n+n_{\rm mod})^2 = (k+1)^2n^2$).  Thus, for
moderate-sized $k$, the pseudo test approximates the power of the
one-sample K-S test, and is faster than both the one-sample and the
two-sample K-S tests for a general model distribution.

\begin{table*}
\begin{minipage}{115mm}
\caption{Comparison of the performance of the pseudo one-sample
2D K-S test and the FF two-sample K-S test as a function of
$k=\frac{n_2}{n_1}$ for a model with $CC=0.10$.} 
\label{tbl_pseudo_vs_real}
\begin{tabular}{@{}rcccccc}
 & & pseudo one-sample test & & & two-sample test \\
$k$ & $n_1$ & $Z_{n,\rm 1s}$$^a$ & $P(>Z_n)$ 
&$\overline{n}$$^b$ & $Z_{n,\rm 2s}$$^a$ & $P(>Z_n)$$^c$ \\
1 & 880 & $1.78\pm 0.07$ & & 440 & $1.30\pm 0.04$ & $34\pm 5\%$ \\ 
2 & 880 & $1.72\pm 0.05$ & & 587 & $1.44\pm 0.03$ & $19\pm 3\%$ \\ 
5 & 880 & $1.56\pm 0.04$ & & 733 & $1.48\pm 0.04$ & $16\pm 4\%$ \\ 
10 & 880 & $1.88\pm 0.05$ & & 800 & $1.79\pm 0.03$ & $3\pm 1\%$ \\ 
20 & 880 & $1.59\pm 0.06$ & $\la 4\%^d$& 838 & $1.64\pm 0.03$ & $8\pm 2\%$ \\ 
50 & 880 & $1.62\pm 0.04$ & $\la 2\%^d$& 863 & $1.65\pm 0.05$ & $8\pm 3\%$\\ 
100 & 880 & $1.62\pm 0.06$ & & 871 & $1.68$$^e$ & $\sim 7\%$\\ 
200 & 880 & $1.66\pm 0.06$ & & 876 & $1.74$$^e$ & $\sim 6\%$\\ 
\end{tabular}

\medskip 
$^a$ Errors determined from K-S comparisons of
10 different subpixelizations of the same simulations.
Systematic uncertainties associated with representing a continuous
model by a discrete distribution are not included.

$^b$ For the two-sample test $\overline{n} = \frac{n_1 n_2}{n_1+n_2} =
\frac{kn_1^2}{n_1+kn_1} = \frac{k}{k+1}n_1$.

$^c$ Obtained from Table~\ref{tbl_cc0.0} for
$\overline{CC} = 0.10 \approx 0.0$ of the two samples.

$^d$ Interpolated from lines 1 ($k\approx 50$) and 2 ($k=25$) of
Table~\ref{tbl_pseudo} ($CC=0.10 \approx 0.12$).

$^e$ Only one comparison was performed due to the longer running time.
\end{minipage}
\end{table*}

\begin{table*}
\begin{minipage}{120mm}
\caption{Critical values $Z_{n,SL}$ for simulated source
distributions. Model size $n_{\rm mod}=45000 \approx 50 \times 880$, 
i.e.\ $k\approx50$.} 
\label{tbl_pseudo}
\begin{tabular}{@{}rrcccccccccc}
& &$SL(\%)$$^\dag$ & 30 & 40 & 50 & 60 & 70 & 80 & 90 & 95 & 99 \\ 
$CC$ &$n$$^\ddag$ & \# of simul. & & & & & & & & & \\ 
0.12 & 880 &10000 &0.98&1.03&1.08&1.13&1.20&1.27&1.38&1.47&1.67\\
0.12 & 1800 &10000 &1.03&1.08&1.13&1.19&1.24&1.32&1.43&1.54&1.74\\ 
0.23 & 820 &10000 &0.95&1.00&1.05&1.09&1.15&1.22&1.33&1.43&1.61\\ 
0.23 & 1800 &10000 &1.02&1.07&1.11&1.17&1.23&1.30&1.42&1.52&1.71\\ 
\end{tabular}

\medskip
$^\dag$ Significance level ($SL \equiv 1-P(>Z_n)$).

$^\ddag$ Size of sample.
\end{minipage}
\end{table*}

Table~\ref{tbl_pseudo_vs_real} presents a comparison of the
performance of the two versions of the test, using the same
five-source setup as in Section~\ref{sec_2d3d}, with one source
shifted by only 0.4~FWHMs~= 4~pix between the model and the simulations.  
As expected,
the power of the two-sample test increases  with increasing $k$, but
for $k\geq20$ the results for $Z_{n,\rm 2s}$ do not change
significantly.  The values of $Z_{n,\rm 1s}$ from the pseudo
one-sample test exhibit greater random variations for small $k$ than
those of $Z_{n,\rm 2s}$.  These are due to the fact that the
small-number statistical uncertainties in the two samples are not
averaged out in the one-sample test, whereas in the two-sample test
they are.  For $k\geq20$ however, the values for $Z_{n,\rm 1s}$ become
self-consistent (as well as consistent with those from the two-sample
test), and there is little power to be gained from increasing $k$.

Despite the similar behaviour of the FF K-S test and our pseudo one-sample
test, the reference tables for the former (Tables~A1--A5 in FF;
Table~\ref{tbl_cc0.0}) cannot be used for the latter, since the $Z_n$
distributions are different.  We therefore ran Monte Carlo simulations
to create a separate look-up table for the new test.
Table~\ref{tbl_pseudo} is based on comparing simulated source
distributions with $n_1 \approx 850$ counts to $k \approx50$ times
brighter simulations (``models,'' $n_2 \approx 45000$) of the same source
distributions.  Sources were simulated on a $128 \times 128$~pix field
using the analytic 5-arcsec (10-pix) HRI PSF.  We also ran Monte Carlo
simulations with higher $n_1$ ($n_1 \approx 1800$) against $n_2
\approx 45000$-count models for the purpose of interpolation.
Although in this case the value of $k$ is lower ($k=25$), we choose to
set the value of $n_2=45000$ as the standard, as it is a measure of
the precision of the Monte Carlo integration.  High values of the
sample correlation coefficient  $CC$ were not pursued,  since
astronomical images rarely have highly correlated photon distributions
(esp.\ in crowded fields, but also given random background).  Thus
constructed, Table~\ref{tbl_pseudo} should be applicable  to most low
background, low $S/N$ imaging cases (esp.\ x-ray).

Using Table~\ref{tbl_pseudo} for the values of $Z_{n,SL}$, the results
for the K-S probability $(P>Z_n)$ from the pseudo one-sample K-S test
will be consistent with those from the FF K-S test using the FF
tables.  Hereafter, unless explicitly stated otherwise, we shall refer
to our pseudo one-sample 2D K-S test as ``the 2D K-S test,'' or simply
as ``the K-S test.''

\subsection{Application of the two-dimensional K-S test to simulated
images \label{sec_sims}}

\begin{figure}
\includegraphics{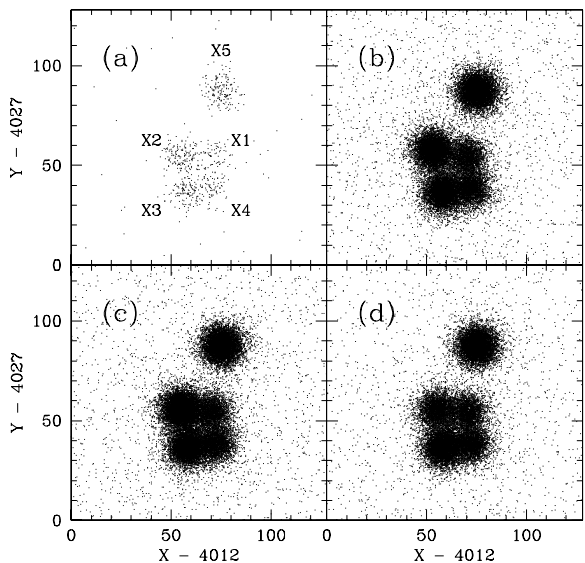}
%\vspace{7.9cm}
\caption{Scenarios for determining the sensitivity of
the pseudo one-sample K-S test with respect to changes in the parent
distribution.  The $x$- and $y$-axes are HRI pixels.  A 10-pixel FWHM
PSF has been used. (a) A
subpixelized 880-count 5-source simulation.  The source parameters are
listed in Table~\ref{tbl_6sou} (X1--X5). The simulations in (b)--(d) are
50 times brighter with a single source modified in each: (b) source X2
is moved 2~pix up and 2~pix to the left; (c) source X2 is 1.45 times
brighter (11600~cts); (d) source X2 is at 60 per cent intensity (4800~cts).}
\label{fig_exper}
\end{figure}

After establishing that the 2D K-S test can be successfully applied to
images, it is important to determine the sensitivity of the test with
respect to deviations from the proposed model.  Here we test the
responsiveness of the 2D K-S test with respect to changes in the
parameters of individual sources in the model distribution.  We look
for the minimum deviation in a single parameter (position or intensity
of a source) that enables the test to tell the distributions apart at
the $\ga 99$~per cent significance level.  As a trial source distribution
we use the one simulated in Figure~\ref{fig_exper}a, whose parameters
are listed in Table~\ref{tbl_6sou} (sources X1--X5), and  compare it to
models that have one parameter changed.

\begin{table}
\caption{Parameters of the source distribution best describing
the x-ray emission in the central $128 \times 128$~pix section of the
NGC~6397 {\it ROSAT} image.} 
\label{tbl_6sou}
\begin{tabular}{@{}ccccc}
Source & Pixel & coordinates & Counts & Counts \\ 
& $(x-4012)$ & $(y-4027)$ & & $(\times 50)$\\
X1 & 71 & 55 & 78 & 3900 \\ 
X2 & 56 & 56 & 176 & 8800 \\ 
X3 & 59 & 37 & 161 & 8050 \\ 
X4 & 73 & 39 & 88 & 4400 \\ 
X5 & 76 & 88 & 197 & 9850 \\ 
X6 & 94 & 77 & 16 & 800 \\ 
bkg & & & 0.015 cts/pix & 0.75 cts/pix\\ 
\end{tabular}
\end{table}

It is worth noting here, that given the significant overlap of PSFs in
a crowded field such as the one in Figure~\ref{fig_exper}a, the
sensitivity of the test with respect to changes in the distribution
need not be isotropic.  It depends on the relative positioning and
brightness of the sources.  We therefore test several possible
scenarios of such changes.  The scenarios shown in
Figure~\ref{fig_exper} and described in its caption correspond to the
limiting cases, in which the K-S test can distinguish the
distributions at the $\ga 99$~per cent level.

After performing K-S tests between the simulation in
Figure~\ref{fig_exper}a and the models in Figures~\ref{fig_exper}b-d,
we establish that the positional sensitivity of the K-S test in
crowded fields depends on the relative source intensity, and on the
direction in which a source is allowed to move with respect to the
crowded region.  The sensitivity to moving a source is higher for brighter 
sources
(3 to 4~pix for source X2) than for the fainter ones (4 to 9~pix for source
X1), and is generally (although not conclusively) poorer when the
source is moved toward the region of crowding (9~pix for source X1,
3~pix for X2) as opposed to when it is moved away (4~pix for X1 and
X2).  The brightness sensitivity of the K-S test is also dependent on
the relative source intensity in a crowded field: brighter sources can
vary by a smaller percentage ($\sim 45$~per cent for X2) than fainter sources
($\sim70$~per cent for X1).  For a source that is sufficiently far away ($\sim
3$~FWHM) from the crowded region (source X5) the sensitivity of the
K-S test is greater and nearly position-independent.  For source X5 we set
the limits at a 2~pix shift in any direction ($P(>Z_n) \approx
1$~per cent) or a 30~per cent change in intensity ($P(>Z_n) < 1$~per cent).

The above-determined sensitivity limits are based on varying 
only one parameter (position or intensity) of a single source, while 
keeping all other parameters fixed.  This is the approach used to
determine the 95~per cent ($\approx 2\sigma$) confidence  
limits (Table~\ref{tbl_coords}) on the positions and
intensities of the detected sources in our NGC~6397 image
(Section~\ref{sec_ngc}).

\begin{table}
\caption{Number of sources in the best-fit model vs.\ K-S
probability.} 
\label{tbl_ngc_sources}
\begin{tabular}{@{}ccccl}
$N$ & $CC$ of model & $Z_n$ & K-S prob $P(>Z_n)$$^a$ &
$P'_N$$^b$ \\
3 & 0.21 & $1.88\pm 0.04$ & $<1\%$ & $1.5\%$ \\ 
4 & 0.22 & $1.48 \pm 0.05$ & $4 \pm 1\%$ & $43\%$ \\ 
5 & 0.19 & $1.23\pm 0.05$ & $22 \pm 6\%$ & $90\%$$^c$ \\ 
6 & 0.21 & $1.19 \pm 0.04$ & $26 \pm 6\%$ & $(100\%)$\\ 
\end{tabular}

\medskip
$^a$ For $n=980$ counts in the NGC~6397 image.

$^b$ Probability that a best-fit model with $N$ sources
and the {\it ROSAT} image represent the same parent distribution; $1-P'_N$
is the significance of adding an $(N+1)$-st source.  $P'_N$ is
calculated as $P_{N,5}$ (Section~\ref{sec_discussion}).

$^c$ Estimated as $P_{5,6}$, i.e.\ the probability that a
5-source best-fit distribution can represent a 6-source one;
$1-P_{5,6} = 10$~per cent is the significance of adding a sixth source.
\end{table}

\section{The 2D K-S test in parameter point estimation for source detection}

\subsection{Algorithm \label{sec_algorithm}}

An iterative source-fitting algorithm was devised that aims
to minimise the $Z_n$ statistic, thus maximising the probability that
an image and a simulation represent the same parent distribution of
sources.  The final simulation that results from this algorithm will
contain the best estimate for the number, positions and intensities of the
sources in the image, subject to limitations arising from the
sensitivity of the test.  The iterative procedure steps
through the following algorithm:
\begin{enumerate}
\item An initial guess of the source distribution is made.  This can 
be a source at the location of the brightest pixel (thus starting
with a one-source configuration) or a guess 
with $N_{\rm initial}>1$ number of sources.
Both initial guesses will produce the same results for a distribution
with $N_{\rm final} \geq N_{\rm initial}$ sources.  The PSF is fit 
to a single unresolved source in an uncrowded part of the image 
so that aspect or other image systematics are included.
\item A bright simulation (a.k.a.\ a ``model''; see
Section~\ref{sec_1s2s}) based on the current guess for the source
distribution is created (using the fitted PSF) and normalised to the image 
intensity.  The
normalised model and the image are smoothed with a Gaussian function
to roughly match the {\it ROSAT}/HRI resolution.  The residual 
between the two is then formed.
\item The next guess for the source distribution is obtained by moving
the source positions against the steepest gradient in the residual,
and by adjusting the intensities, so as to decrease the
maximum deviation from zero ($D_{\rm max}$) in the smoothed residual.
We indeed observe that $D_{\rm max}$ is strongly correlated to the value of
$Z_n$ (correlation is 0.93; Figure~\ref{fig_zn_res}), and hence (for
constant $CC$) to the K-S probability $P(>Z_n)$.

\begin{figure}
\includegraphics{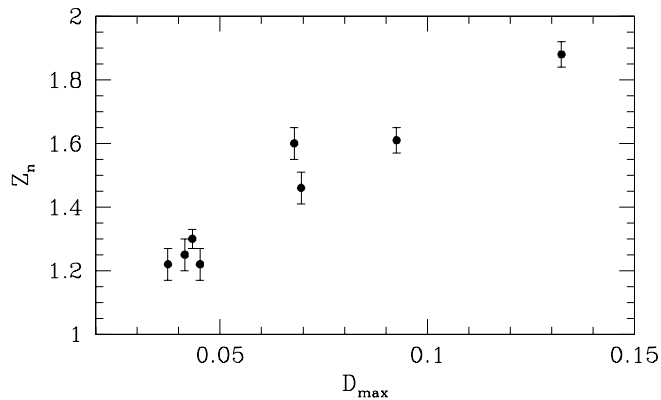}
%\vspace{5.8cm}
\caption{Values of $Z_n$ vs.\ the maximum deviation
from zero, $D_{\rm max}$, in the smoothed residual.  The error-bars in
$Z_n$ represent the standard deviation of $Z_n$ due to 10 independent
subpixelizations of the compared simulations.  The statistics $Z_n$
and $D_{\rm max}$ are highly correlated (correlation coefficient
0.93), and we therefore use a minimum in $D_{\rm max}$ as an 
indication of being near a minimum in $Z_n$ (and hence, for constant
$CC$, near a maximum in $P(>Z_n)$).}
\label{fig_zn_res}
\end{figure}

\item Repeat steps (ii) and (iii) until $D_{\rm max}$ is
minimised. Compare the image to the model with the 2D K-S test and, 
if necessary, further minimise $Z_n$ by applying small changes (e.g.\
single-pixel shifts) to the model (since the nature of the relation
between $D_{\rm max}$ and $Z_n$ is not established rigorously).  The
final simulation will contain the best guess of the positions and
intensities of the assumed sources.
The image is compared to the model using the 2D K-S test.
\item Steps (i) through (iv) are run for a fixed number of sources
(guessed in step (i)).  If the final K-S probability of similarity is
not satisfactory (e.g.\ not $\ga 5$~per cent), a new source is added at the
location of the highest residual, and the algorithm is repeated from
step (ii).
\item If the addition of the last source did not incur a decrease 
in the $Z_n$ statistic larger than its uncertainty 
($\pm 2$~per cent to $\pm 4$~per cent), the last added source is considered
marginal, and the previous best guess for the number, positions and
intensities is taken as the final one.
\end{enumerate}

As noted in step (i) an initial guess with $N_{\rm initial} > 1$
number of sources can also be fed into the algorithm.  Such a guess
can be made either from visual inspection of the 
image, or after applying a deconvolution algorithm.  We found that
L-R deconvolution (see Section~\ref{sec_compare} below) gives
good initial estimates of the positions of the individual
sources.  However, since deconvolution can introduce spurious sources,
the initial guess of the number of sources should be conservative.

\subsection{Performance \label{sec_performance}}

The above procedure has not been automated, and therefore (due to
subjectivity in ``guessing'' a simulated source distribution after
having created it) we have not performed tests to explicitly determine
its efficiency in detecting sources.  We quote the ability of the
2D K-S test to detect small changes in the positions (within $\sim
0.2$~FWHM) and/or
intensities of individual sources (Section~\ref{sec_sims}) as an
indication of the power of the iterative algorithm.  Nevertheless, we
have devised a method to test the confidence with which a certain
number of sources can be claimed in a given photon distribution.  The
method takes our best guess for the source distribution in the image
with a given number of sources (e.g.\ $N$), and compares a model of it
to a faint simulation of our best guess with one source fewer ($N-1$).
In this way we can test in what fraction $P_{N-1,N}$ of the cases our
proposed model (with $N$ sources) can describe a source
distribution with $N-1$ sources.  In other words, we test for the
significance ($1-P_{N-1,N}$) of the addition the $N$-th source;
$P_{N-1,N}$ is thus its false-detection probability.  If this
comparison is performed many of times (of the order of the
number of Monte Carlo simulations done for each row in
Table~\ref{tbl_pseudo}), a $Z_n^{N-1,N}$ curve
for the two guesses is recovered.  The latter can be then compared to a
$Z_n^{N,N}$ curve, obtained in a similar fashion comparing $N$-source
simulations to an $N$-source model.  

\begin{figure}
\includegraphics{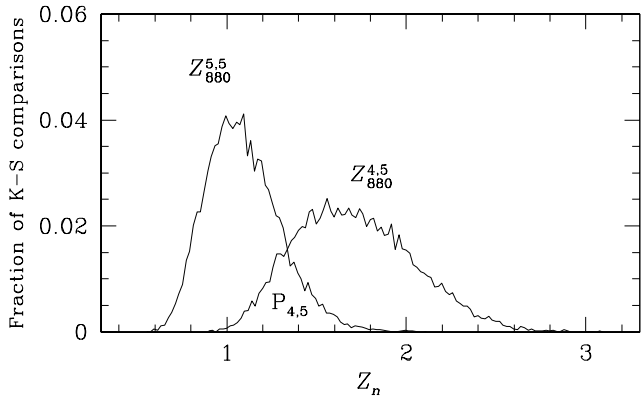}
%\vspace{5.7cm}
\caption{ A comparison between a $Z_{880}^{5,5}$
and a $Z_{880}^{4,5}$ curve obtained using the 5-arcsec predicted
HRI PSF.  The $Z_{880}^{4,5}$ curve is obtained from K-S
tests between a 5-source model and 10000 realisations of a 4-source
simulation that best represents the 5-source model.  The overlap
area $P_{4,5}$ determines the false detection probability of the 5th
source.  Here $P_{4,5}=0.23$.} 
\label{fig_zn_overlap}
\end{figure}

Example $Z_n^{N,N}$ and $Z_n^{N-1,N}$ curves for $N=5$ and $n=880$ 
are shown in Figure~\ref{fig_zn_overlap}.  
The overlap of the two curves gives the fraction of Monte
Carlo simulations, in which a best-fit distribution with $N-1$ sources
produces an image that has the same K-S similarity to the model as
that of a best-fit distribution with $N$ sources.  The ratio of the
overlap area to the area of either of the $Z_n$ curves
(assuming they are both normalised to the same area) is the desired
fraction $P_{N-1,N}$.

To investigate the dependence of the overlap area $P_{N-1,N}$ on the
width of the PSF, we ran K-S tests simulations built with an 8-arcsec
Gaussian PSF.  The result is that for a wider PSF the
$Z_n^{N-1,N}$ curve is narrower and is shifted toward smaller $Z_n$
(the model and the simulations look more alike).
The $Z_n^{N,N}$ curve however is not affected.  The overall effect is
that the false-detection probability $P_{N-1,N}$ of the $N$-th
source increases (from 0.23 to 0.43 for $N=5$; Table~\ref{tbl_ngc_sources}).

The above technique can be generalised to produce $P_{N-i,N}$ for
arbitrary integers $i<N$ and $j$.  An application of
$P_{N-i,N}$ is discussed in Section~\ref{sec_discussion}.

\section{Application to a deep {\it ROSAT} image of NGC~6397} 
\label{sec_ngc}

\begin{figure}
\includegraphics{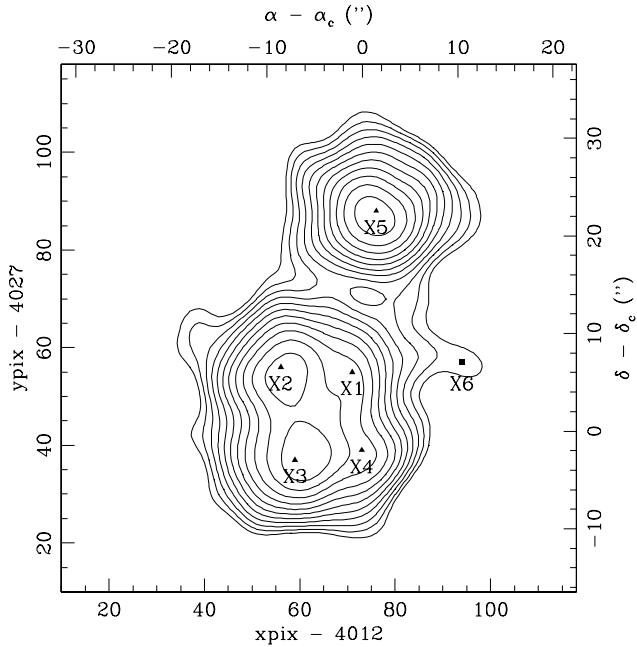}
%\vspace{9.1cm}
\caption{A 75~ksec {\it ROSAT}/HRI exposure of the central
region of NGC~6397, with the detected sources marked.  The x-ray
image is smoothed with a 2-d $\sigma=2$ arcsec Gaussian.  The cluster
centre is at $(\alpha_c, \delta_c) =$ (17:40:41.3, --53:40:25),
\citep{djo93}. The conversion from pixel to celestial coordinates is
accurate to within 1~arcsec.}
\label{fig_ngc}
\end{figure}

The iterative source-modelling procedure (Section~\ref{sec_algorithm})
was applied to our 75~ksec {\it ROSAT}/HRI exposure (March 1995) 
of the core region of
the post core-collapse globular cluster NGC~6397
(Figure~\ref{fig_ngc}).  Standard aspect correction routines
\citep{har98a, har98b, har99a, har99b} were applied to the image to
improve the $S/N$ ratio.  After the aspect corrections the
PSF improved from 10.3~arcsec $\times$ 8.3~arcsec to 8.3~arcsec
$\times$ 7.9~arcsec, as 
measured from the shape of a background point-source quasi-stellar
object (QSO) at 3.7~arcmin off-axis \citep[source ``D'' in][]{coo93}.  
The obtained size of the PSF was still much worse 
than the predicted 5~arcsec.  This effect is not due to the known 
deterioration of the {\it ROSAT} PSF with increasing off-axis angle beyond 
$\sim 5$~arcmin, since the QSO is only 3.7~arcmin from the centre
of the field. Residual (unknown) aspect errors are present, and
in the analysis below we use a fitted PSF instead of the nominal one.

We analyse the central 128 pixel (64-arcsec) square region of the
image, containing 980 counts.  Model simulations were created using an
analytical PSF fit to the QSO with the {\sc iraf/daophot} routines
{\sc psf} and {\sc addstar}.  The PSF was
comprised of a FWHM $\approx 8$-arcsec Gaussian core and Lorentzian wings,
where the core and the wings could be tilted along different
directions in the image.  In determining
the false-detection probabilities (from the overlap of the $Z_n$
curves) however, for faster iteration we used a symmetrical 8-arcsec 
Gaussian PSF, noting that the $Z_n$ distributions based on the fitted PSF 
and on the Gaussian PSF are expected to be indistinguishable, since the
test is distribution-free.

Table~\ref{tbl_ngc_sources} presents results for the K-S statistics of
the best-fit models for a given number of sources.  The error in the
values of $Z_n$ and $P(>Z_n)$ is the one-sigma uncertainty due to
subpixelization, as determined from K-S comparisons between the model
and 10 independent subpixelizations of the same image.  It is an
estimate of the error in the mean of the $Z_n$ distribution of
comparisons between the image and the $N$-source model.

\begin{figure}
\includegraphics{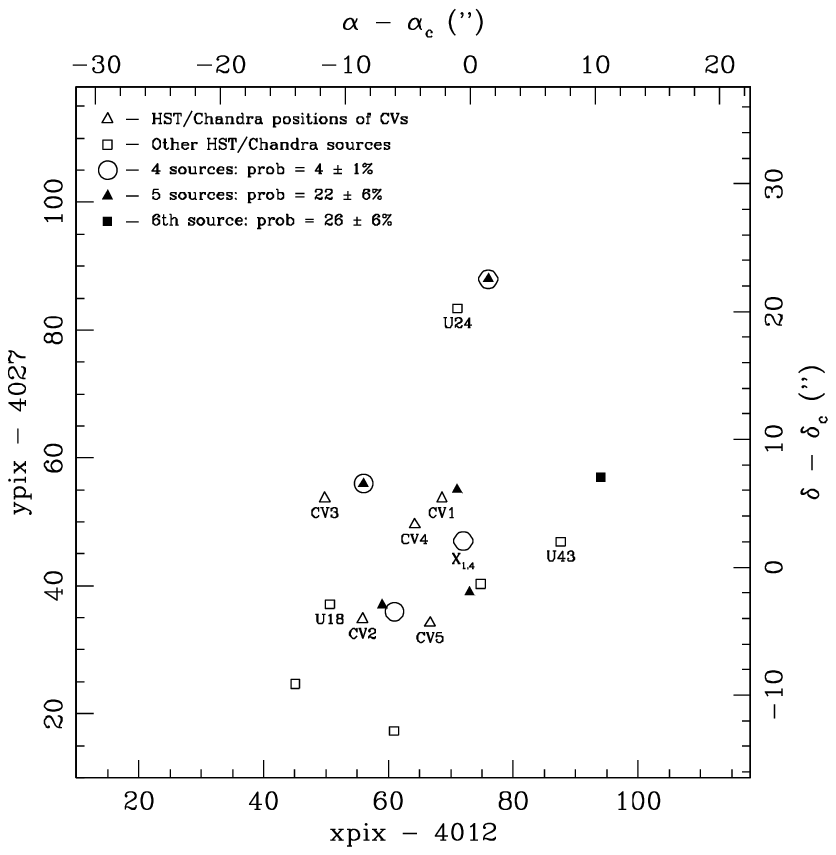}
%\vspace{9.1cm}
\caption{Best-fit source positions for 4, 5 and 6 sources.
Open triangles mark the positions of known H~$\alpha$-emission 
objects \citep[probable CVs, numbered with their ID from][]{coo95,
gri99}~-- candidate counterparts for sources X1--X4.  Open squares
mark $UV$-excess \citep{coo95, edm99} and/or faint
($L_x<10^{31}$~erg~s$^{-1}$) {\it Chandra} sources
\citep[IDs from Fig.~1 of][]{gri01a}.
The consistent shift between the {\it HST}/{\it Chandra} CV and the
{\it ROSAT} x-ray 
positions is indicative of a bore-sight offset ($\sim 2$~arcsec) of {\it ROSAT}
for this observation.} 
\label{fig_ngc_sources}
\end{figure}

Following the logic of step 6.\ in the K-S probability maximisation
algorithm (Section~\ref{sec_algorithm}), we conclude that four sources
are insufficient to represent the image conclusively, since the
addition of a fifth source decreases the $Z_n$ statistic by more
(17~per cent)
than the 3.5~per cent error in the $Z_n$ statistic for four
sources.  However, the addition of a sixth source is not justified,
since the decrease (0.04) in $Z_n$ is smaller than the
error (0.05). 
We therefore claim that five sources are sufficient, and that at
least four sources are necessary (at the $1-P_{3,5} > 99$~per cent level) to
account for the observed photon distribution in the {\it ROSAT} image.
The source centroids for the optimal source configurations
with 4, 5 and 6 sources are shown in Figure~\ref{fig_ngc_sources}.
The number of counts per source for the five- and six-source cases are
the same as listed in Table~\ref{tbl_6sou}.

Source X6 is sufficiently faint and detached from the central
group (X1--X4), that its addition did not necessitate any changes in
the prior five-source configuration.  It is 11~arcsec away from the
closest source (X1), and 4.5 times dimmer than the faintest one (also X1; 
Table~\ref{tbl_6sou}).  The false detection probability
(determined as $P_{5,6}$) for source X6 is 90~per cent.  

Source X$_{1,4}$ in the four-source best-fit solution falls approximately in
the middle between sources X1 and X4 of the five-source solution,
which is consistent with them having comparable intensities and being
fainter than X2 and X3 (Table~\ref{tbl_6sou}). 

The derived optimal number of five sources in our 75~ksec image of NGC~6397 is
consistent with the earlier suggestion \citep{coo93} that at least
four x-ray sources (X1, X2, X3 and X5) are present in an 18~ksec exposure
of the same region (found by visual inspection of the peaks in the
image, and confirmed by a one-dimensional azimuthal K-S test around
the source centroids).  The locations of the detected sources are also
consistent (up to a $\sim 2$-arcsec systematic offset) with the
positions of known optical cataclysmic variables (CVs): the optimum
five-source positions (solid triangles, Figure~\ref{fig_ngc_sources})
are systematically displaced by $\sim 2$~arcsec to the lower right of
CV1--CV5 (open triangles, as measured from H$\alpha$, $\cal R$, and
$UBVI$ {\it HST} data; \citealt{coo95, coo98}).  Such an offset is well
within the expected variation ($\la 5$~arcsec) of the
absolute pointing of {\it ROSAT}.

Source X5 does not have a known optical counterpart, and sources CV1
and CV4 are unresolved in the {\it ROSAT} x-ray image due to their
proximity ($\sim 2.5$~arcsec $\approx 0.3$~FWHM), so that the emission
of X1 is probably due 
to both of them.  The object CV5 was previously known \citep{coo98}
as a near $UV$-excess star, but proposed as a CV-candidate
\citep{gri99, met99} only after re-evaluating its probable association with
the x-ray emission from X4.  Its subsequent confirmation as a
CV-candidate (H$\alpha$-emission object) in our followup deep {\it HST}
imaging \citep{tay01} is indicative of the reliability of our
iteration scheme to detect faint sources in crowded fields.

Similarly, source X6 (if real) may be associated with the {\it
Chandra} source U43 \citep{gri01a}: a probable faint
BY~Draconis binary, identified as PC-4 in \citet{tay01}.  
Such objects are expected in globulars as the binary 
companions of stars on to which they have transferred their envelopes,
and indeed the reported velocity in \citet{edm99} suggests a massive
but dark binary companion.  The latter is most likely a neutron star
(NS), since He-WD/NS systems are expected in millisecond pulsars
(MSPs), and MSPs are detected as faint x-ray sources with luminosities
comparable to that of X6 \citep[cf.][]{bec99}.

Table~\ref{tbl_coords} lists results for the detected sources.  A mean
bore-sight offset of $-3.5 \pm 1.0$~pix in $x$, and $+1.7 \pm 1.0$~pix
in $y$ (--1.8~arcsec and +0.85~arcsec, respectively) has been applied to the
x-ray positions to align them with the suggested optical counterparts.
In calculating the offset, we discard source X1 (since the x-ray
emission in its vicinity comes most likely from both CV1 and CV4), and
weigh the measured shifts inversely to the square of the positional
uncertainties of the sources along $x$ and $y$.  The 95~per cent
confidence radius 
of the source coordinates corresponds to the maximum shift of a single
source from its listed position (keeping all intensities and other source 
positions constant) that maintains the K-S probability above 5~per cent.

\begin{table*}
\begin{minipage}{163mm}
\caption{X-ray sources detected in NGC~6397.} 
\label{tbl_coords}
\begin{tabular}{@{}lcccccccc}
Source & R.A. & DEC. & 95\% conf. & Count rate & $L_x$ (erg/s)$^a$ &
Optical & \multicolumn{2}{c}{{\it Chandra} offset$^d$ (arcsec)}\\
 & (J2000) & (J2000) & radius & (ksec$^{-1}$) & $kT=10$~keV & 
counterpart$^c$ & $\alpha_R-\alpha_C$ & $\delta_R-\delta_C$\\ 
X1 &17:40:41.46 &$-$53:40:18.4 &$2.5\arcsec$ &$1.2\pm0.5$ &$3.3 \times
10^{31}$ &CV1, CV4 &$-$0.2 &$-$0.6\\
X2 &17:40:42.48 &$-$53:40:18.0 &$1.5\arcsec$ &$2.7\pm0.6$ &$7.4 \times
10^{31}$ &CV3 &0.7 &1.2\\
X3 &17:40:42.32 &$-$53:40:27.4 &$1.5\arcsec \times 2.5\arcsec$ &$2.5\pm0.6$
&$6.9 \times 10^{31}$ &U18, CV2 &0.6 & $-$1.8\\ 
X4 &17:40:41.53 &$-$53:40:26.4 &$2.0\arcsec$ &$1.3\pm0.5$ &$3.6 \times
10^{31}$ &CV5 &0.7 &$-$0.1\\
X5 &17:40:41.36 &$-$53:40:02.0 &$1.2\arcsec$ &$3.0\pm0.6$ &$8.2 \times
10^{31}$ &U24 &0.0 & 0.0\\
X6 &17:40:40.4 &$-$53:40:18 &$\infty^b$ &$0.4$ &$<1.1 \times 10^{31}$
&U43 &0.8 &2.8\\
\end{tabular}

\medskip
$^a$ Unabsorbed luminosities listed in the 0.5--2.5~keV
band,  for a cluster distance of 2.2~kpc, and column density of $1.0
\times 10^{21}$~cm$^{-2}$.  For best fit ({\it Chandra}) column
densities and bremsstrahlung spectra for individual sources, see
\citet{gri01a}.

$^b$ The $2\sigma$ confidence radius of the position of X6 is infinite,
because the K-S probability that the model and the image 
represent the same parent distribution is always above 5 per cent,
regardless of the source location.

$^c$ U18, U24 and U43 are \citet{gri01a} {\it Chandra} IDs. U18 also
identified as either a BY~Dra or MSP by \citet{gri01a}, and U43
identified as a BY~Dra binary by \citet{tay01}. CV2 first identified
as H$\alpha$ object by \citet{coo95}.

$^d$ Offset between the given positions (subscript $R$) and the ones
listed in \citet[][subscript $C$]{gri01a}. A boresight offset
($\Delta\alpha_{R-C}=4.9$~pix and $\Delta\delta_{R-C}=4.6$~pix) has been
applied to match the positions of our
best-constrained source (X5) and its {\it Chandra} counterpart (U24).
When the emission from two {\it Chandra/HST} sources
corresponds to a single {\it ROSAT} source, the latter has been
associated with the mean position of the {\it Chandra/HST} sources.
\end{minipage}
\end{table*}

\section{Comparison to other source detection methods} 
\label{sec_compare}

To get an idea of the superior performance of our source-modelling
scheme we compared it to established source-detection algorithms, such as
the classical ``sliding-cell'' detect, the wavelet detect,
the {\sc iraf/daophot} PSF-fitting task {\sc allstar}, and ML analysis.  We also used
deconvolution routines on the image to determine possible source
locations.  Below we discuss briefly each  of these alternatives.

{\it Sliding-cell Detect:}  The sliding-cell detect algorithm is based
on $S/N$ calculation and was not expected to perform well in a crowded
low-$S/N$ field.  Indeed, the two versions of this algorithm in the
{\sc iraf/pros} package (tasks {\sc imdetect} and {\sc ldetect} in the
{\sc xspatial} package)
fail to produce the expected number of x-ray sources in the cluster.
{\sc imdetect} uses a constant average background for the entire image, and
a variable detect cell size (squares with sides from 4~arcsec to 24~arcsec)
to search for sources.  The larger detect cells fail to find more than
three sources in the central region of NGC~6397, whereas the 4-arcsec
detect cell size is too small for use with our PSF (FWHM $\approx
8$~arcsec), and produces an unjustified high number of individual sources.
The local detect algorithm (task {\sc ldetect}) calculates $S/N$ around each
pixel, using the local background (in a region between 1.5~arcsec and
2.5~arcsec from the source) as an estimate of the noise.  As a result it
does not handle crowded fields adequately, and cannot distinguish
blended sources.  Even the two smallest detection cells (6~arcsec $\times$
6~arcsec and 9~arcsec $\times$ 9~arcsec) do not find more than 3
sources in the image in Figure~\ref{fig_ngc}.

{\it Wavelet Detect:} This algorithm based on the wavelet transform
has only recently been applied to imaging astronomy \citep[][and references
therein]{fre01, dam97}, and has been demonstrated to outperform other
source detection algorithms in low-$S/N$ fields.  We used an
implementation of the wavelet detect based on the Marr wavelet, or the
``Mexican Hat'' function, coded in the {\sc wavdetect} task in the
{\it Chandra} {\sc detect} 1.0 Package.  The algorithm
is most sensitive to structures of size approximately equal to the
width of the Mexican Hat function.  Running {\sc wavdetect} on our NGC~6397
image (Figure~\ref{fig_ngc}) with transforms of width $\leq 8$~arcsec
produced only the same three sources already found by the sliding-cell
algorithms.  This was not unexpected, since in simulated images for
the {\it Chandra} High Resolution Camera (FWHM = 0.5~arcsec), {\sc
wavdetect} is unable to discern point sources less than 2 FWHM
apart.\footnote{{\it Chandra} {\sc detect} User's Guide; URL:
http://hea-www.harvard.edu/asclocal/user/swdocs/detect/html/}

{\it Image Deconvolution:} There exist a number of widely used image
deconvolution algorithms that are applicable to moderately crowded
fields.  After comparing results from the {\sc iraf} implementations of the
Maximum Entropy Method, the L-R algorithm (both applicable primarily to
optical images), and from {\sc clean} (used mostly in radio imaging), we
found that L-R deconvolution \citep{luc74, ric72} most reliably discerns
the five-source distribution found by our iterative
source-modelling scheme (Section~\ref{sec_ngc}).  The positions of the
peaks in the deconvolved image are in excellent agreement (to within
$\pm 1$~pix = $\pm 0.5$~arcsec) with
the K-S best-fit source positions, which
exemplifies the usefulness of L-R deconvolution in analysing crowded
fields.  Unfortunately, the L-R method does not provide a measure of
the goodness of fit of these positions and of the significance of the
peaks in the reconstructed image.  These need to be determined
separately with a multi-source fitting routine (since the field is
crowded), such as {\sc daophot/allstar}, or the current (2D K-S) iterative
method.  Furthermore, the obtained intensities of the deconvolved
sources are in 
much poorer agreement with the ones from the 2D K-S best-fit model.
Nevertheless, L-R deconvolution does give an indication for the
existence of more than 3 sources (which could not be determined with
the source-searching methods).  The L-R method thus provides a very
good initial guess for the source configuration, which can be input to
iterative source-modelling algorithms.

{\it {\sc daophot/allstar:}} The {\sc daophot} package is designed for the
analysis of crowded optical images, and as such it assumes that the
images are in the Gaussian statistics (high number of counts per
pixel) regime.  Thus, strictly speaking, the package is inapplicable
to data governed by Poisson statistics, such as most x-ray images
(including our NGC~6397 image,  containing $\leq 3$ counts per pixel),
because it severely  underestimates random errors.  However, until recently
{\sc daophot} was the only widely available software for reduction
of crowded fields, and it has been suggested \citep{coo93} that it
can be useful for analysing crowded x-ray fields.

Our experience with {\sc allstar} is that it is heavily dependent on
several loosely defined parameters which, in regimes of severe source
confusion and low signal-to-noise as in our NGC~6397 image
(Figure~\ref{fig_ngc}), critically determine the
performance of the task.  We found that different combinations of the
values of the parameters and of the initial guess for the source
distribution produced different final results, in which the number of
detected sources in the NGC~6397 image varied from 2 to 5.  
By judiciously adjusting its parameters, {\sc allstar} can be made to 
detect 5 sources, however that combination is not favoured statistically 
over other combinations with fewer sources.  In the
case when {\sc allstar} detects 5 sources, the obtained positions and
intensities are such that the
K-S probability of similarity with the {\it ROSAT} image is $<1$ per cent
($Z_n = 2.2$).

{\it Maximum Likelihood:} Given our method of optimisation -- minimising
the maximum residual $D_{\rm max}$ (albeit we then further minimise the
K-S statistic $Z_n$) -- ML analysis would be expected
to produce a similar fit.  This is indeed the approach of \citet{ver00} in
analysing the same {\sl ROSAT} field.  The results for the 5 detected
sources (Model~I in \citep{ver00}; X1--X5 in this paper) agree well;
in addition, our analysis suggests the possible presence of the
faint source X6.   We choose to employ a 2D K-S test to assess the
goodness of fit instead, banking on its sensitivity to diffuse
distributions.  As pointed out by the referee, it is a good test for the
location of smeared objects, but it is rather 
insensitive to their width.  Via K-S, a source may be deduced to be
unresolved, despite having broader profile, which can
frequently be the case in Poisson noise limited images.

We have thus demonstrated that under conditions of severe source
confusion and low $S/N$, our source-detection method based on a 2D K-S
test works better than other available techniques.  We
attribute its performance to the fact that our approach uses the actual
PSF in searching for sources (sliding-cell and wavelet detect algorithms
do not), that no information is lost to binning (as in the
Pearson $\chi^2$ test, used in {\sc allstar}), and that it
is more sensitive to broad emission than other tests (e.g., ML).

We have not made a comparison of our method against the Pixon
deconvolution method \citep{pin93}.  Our method was originally intended
to enhance sensitivity for crowded point source detection; the Pixon
method also shows good results for the detection of low surface
brightness features.

\section{Discussion}

\subsection{Applicability to distributions with unknown parameters}

Rigorously, the presented look-up Table~\ref{tbl_pseudo} (generated by
comparing simulations of models with {\sl a priori} known
parameters) is not applicable when comparing an
image of an unknown source distribution to a
simulation with known parameters.  \citet{lil67} investigates this
situation for the case of the 1D K-S test and sampling from a
distribution with unknown mean and variance (``the Lilliefors test for
Normality'').  He finds that the standard 1D K-S test table is {\sl too
conservative}, i.e.\ with an appropriately 
generated look-up table (via Monte Carlo simulations), one can reject
the null hypothesis that a sampled distribution is Normal at a higher
significance level than with the standard table.

The implications of this to our case are not known, and have
not been investigated.  Speculatively extrapolating Lilliefors's
conclusion, the 2D K-S test for comparing an unknown to a known
distribution should be, if anything, 
more powerful than presented.  This would increase the significance of
source X6, making its association with the suggested BY~Dra variable
more likely.

The advantage of an ML approach here would be that likelihood ratios
between different models do not suffer from such problems.

\subsection{Significance of the detections} 
\label{sec_discussion}

The developed detection significance test for additional sources in
Section~\ref{sec_performance} may seem subjective, since prior
knowledge is needed about the $Z_n^{M,M}$ curve (where $M$ is the
number of sources in the image).  Naturally, this information is not
available when working with an astronomical image representing an
unknown source distribution, where $M$ is a sought parameter.
However, due the (nearly) distribution-free character of the pseudo
one-sample 2D K-S test (Section~\ref{sec_1s2s}), all that is needed is
the correlation coefficient $CC$ of the counts in the image, which is
readily available ($CC=0.17$ for the {\it ROSAT} image in
Figure~\ref{fig_ngc}).  Provided that the best-fit model with $N$
sources represents the image reasonably well (K-S probability $\ga
1$~per cent), the $Z_n^{N,N}$ curve will be indistinguishable from the
$Z_n^{M,M}$ curve of the image, since the correlation
coefficients of the $N$-source model and of the ($M$-source) image
will be very similar. Indeed, in  our case the best-fit five-source
model has $CC=0.19$, which given the slow dependence of the $Z_n$
distribution on $CC$, well approximates the $Z_n$ distribution for
$CC=0.17$ (the correlation coefficient of the counts in the {\it ROSAT}
image).

More general than the false-detection probability is the
fraction $P'_K$ of cases, in which the observed
image can be represented by a best-fitting model containing $K<N \leq
M$ sources, where $K$ does not necessarily equal $M-1$.  Here $N$ is
our best guess for the number of sources in the image, and $M$ is the
actual (unknown) number of sources.  The quantity $1-P'_K$ is the
significance level at which we can reject the hypothesis that the
image contains only $K$ sources.  To determine $P'_K$ using the
2D K-S test, we need to compare multiple images of the same
field to a single model simulation with $K$ sources
(Section~\ref{sec_1s2s}).  Naturally, this cannot be done, since there
rarely exist multiple available images of the same field.  However,
$P'_K$ is well approximated by the quantity $P_{K,N}$, given
that, as discussed above, $Z_n^{N,N}$ describes well the $Z_n^{M,M}$
distribution of the image.  This is the value
listed in Table~\ref{tbl_ngc_sources} (using $N=5$) for the
probability that the {\it ROSAT} image can be fitted with fewer than 5
sources. For $K=N-1$, in $P_{K,N}$ we recover the false-detection
probability for the $N$th source, as already discussed in
Section~\ref{sec_performance}.

\subsection{Detected sources}

The positions and the count rates of the detected sources are in
excellent agreement with Model~I (based on 1995 {\it ROSAT} data) of
\citet{ver00} in 
their maximum-likelihood analysis of the same HRI field.
A {\it Chandra} image of the core region of NGC~6793 \citep{gri01a}
reveals a greater complexity of sources
(Figure~\ref{fig_ngc_sources}).  The source ``doubles'' CV1 and CV4,
as well as CV2 and U18 are too close ($\sim 2.5$~arcsec $\approx 0.3$~FWHM)
to be distinguished as separate sources in the {\it ROSAT}/HRI image, and
are represented as blended sources X1 and X4, respectively.  The
remainder of the sources marked with open squares are too faint
\citep[$L_x<10^{31}$~erg~s$^{-1}$][]{gri01a} to be detected given the
crowdedness of the field.  None the less, there is a
clear one-to-one correspondence between the brightest ($L_x >
10^{31}$~erg~s$^{-1}$) {\it Chandra} sources, and the ones detected in the
{\it ROSAT}/HRI image using the 2D K-S technique.  

Although in their Model~IV
\citeauthor{ver00} predict the existence of separate x-ray
counterparts to sources CV2 and U18, that model is fit to 1991
{\it ROSAT}/HRI data when CV2 was more prominent in x-rays relative to U18
\citep[hence could be more accurately centroided; cf.\ Fig.~1
in][]{coo93}, and three of the sources in the model have fixed
positions.  On the other hand, in our 2D K-S test iterative analysis
we have not used any fixed parameters.  Moreover, the K-S test
suggests the existence of source U43, detected (albeit inconclusively, and
offset by $\sim 4.5$~arcsec from its {\it Chandra} position) as source X6,
for which there exists no x-ray identification prior to the {\it Chandra}
results of \citet{gri01a}.  Although the source is fainter
\citep[log$L_x=29.4$][]{gri01a} than other undetected sources in the
complex X1--X4, 
the source must have been $\ga10\times$ brighter to have been detectable 
with {\it ROSAT} \citep[indeed, BY~Dra binaries were discovered in 
globular clusters as faint and flaring x-ray sources,][]{gri01b}.

\section{Conclusion}

We have developed an application of the 2D K-S test \citep{pea83, fas87}
in a source-detection algorithm for astronomical images.  By employing
the ``subpixelization'' technique on 3D astronomical images, we show
that the 2D K-S test has greater power than the 3D
K-S test -- the intuitive choice for such images.  We use Monte
Carlo integration to determine the cumulative values of the proposed
model distribution in all four quadrants around each count and,
recognising the deviations that this incurs from the derived $Z_n$
distributions, we provide our own reference tables for estimating the
K-S probability.

We devise an iterative source-modelling routine that employs the K-S
probability as a goodness of fit estimator, and can be used to find
the optimum number, positions, and intensities of blended sources.  We
then apply the iteration scheme to a deep (75~ksec) {\it ROSAT}/HRI exposure
of the core region of NGC~6397 and find five blended sources, as well as a
possible sixth one.  The locations of the five brightest (and possible
sixth) x-ray sources
match closely (within the positional error bars) the locations of 
probable CVs and BY~Dra systems, discovered
with {\it HST} \citep{coo93, coo95, tay01}, and 
confirmed with {\it Chandra} \citep{gri01a}.
The sixth source, X6, is a marginal detection with the 2D K-S 
technique and is likely identified with the much fainter 
Chandra source, U43 \citep{gri01a}, which is in turn 
identified with a BY~Dra binary, PC-4 \citep{tay01}.

Comparisons to other source-detection schemes (sliding-cell, wavelet
detect, {\sc daophot/allstar}, L-R deconvolution and ML
techniques) applied to the same 
image demonstrate the superior power of our method in heavily crowded
fields with low signal-to-noise.  The example with the {\it ROSAT}/HRI deep
field indicates that the proposed iterative source-modelling scheme can
find applications in small-number statistics high-energy imaging,
e.g.\ in deep exposures of globular clusters and extragalactic nuclear
regions with {\it Chandra}, where the size of the 
PSF is often comparable to the angular separation between the objects.

\medskip
We thank our collaborators Adrienne Cool and Peter
Edmonds for numerous discussions.  This research was partially
supported  by NASA/LTSA grant NAG5-3256 and by {\it HST} grant GO-06742.

\end{document}